\documentclass[twocolumn,runningheads,natbib]{svjour2}
\smartqed  % flush right qed marks, e.g. at end of proof
\usepackage{graphicx}
\journalname{Astrophysics and Space Science}

\newcommand{\sub}{{\sl Subaru}}
\newcommand{\hst}{{\sl HST}}

\newcommand{\fors}{{\sl FORS1}}
\newcommand{\isaac}{{\sl ISAAC}}
\newcommand{\vlt}{{\sl VLT}}
\newcommand{\ntt}{{\sl NTT}}
\newcommand{\cxo}{{\sl CXO}}
\newcommand{\xmm}{{\sl XMM}}
\newcommand{\naco}{{\sl NACO}}

\newcommand{\asca}{{\sl ASCA}}
\newcommand{\rosat}{{\sl ROSAT}}
\newcommand{\sax}{{\sl BeppoSax}}
\newcommand{\hstn}{{\sl Hubble Space Telescope}}

\newcommand{\spitzer}{{\sl Spitzer}}
\newcommand{\acsn}{{\sl Advanced Camera for Surveys}}
\newcommand{\acs}{{\sl ACS}}
\newcommand{\stisn}{{\sl Space Telescope Imaging Spectrometer}}
\newcommand{\stis}{{\sl STIS}}
\newcommand{\nuv}{{\sl NUV-MAMA}}
\newcommand{\eclipse}{{\sl eclipse}}

\begin{document}

\title{Studies of Neutron Stars at Optical/IR Wavelengths
\thanks{Based on observations with the NASA/ESA \hstn, 
obtained at the Space Telescope Science Institute, which is operated by
AURA,Inc.\ under contract No NAS 5-26555. }\thanks{Based on observations collected at the European Southern Observatory, Paranal, Chile under  programme ID 63.P-0002(A), 71.C-0189(A)72.C-0051(A),074.C-0596(A), 074.D-0729(A), 075.D-0333(A), 076.D-0613(A)}
}

%\subtitle{Do you have a subtitle?\\ If so, write it here}

%\titlerunning{Short form of title}        % if too long for running head

\author{R. P. Mignani       
\and
S. Bagnulo
\and
A. De Luca
\and
G. L. Israel
\and
G. Lo Curto
\and
C. Motch
\and
R. Perna
\and
N. Rea
\and
R. Turolla
\and
S. Zane
}

\authorrunning{R. P. Mignani et al.} % if too long for running head

\institute{R. P. Mignani \and S. Zane \at
              MSSL-UCL, Holmbury St. Mary, Dorking, RH56NT, UK \\
              Tel.: +44-1483-204267\\
              Fax:  +44-1483-278312\\
              \email{rm2@mssl.ucl.ac.uk}           %  \\
%             \emph{Present address:} of F. Author  %  if needed
       \and
       S. Bagnulo, G. Lo Curto \at
       ESO, A. de Cordova 3107, Vitacura, Santiago, 19001, Chile
       \and
       A. De Luca \at
       INAF-IASF, Via Bassini 15, 20133, Milan, Italy
       \and
       G.L. Israel \at
       INAF-OAR, Via di Frascati 33, 00040, Monte Porzio, Italy
       \and 
       C. Motch \at
       OAS, rue de l'Université 11, 67000, Strasbourg, France 
       \and
       R. Perna \at
       University of Colorado, 440 UCB, Boulder, 80309, USA 
       \and
       N. Rea \at
       SRON,Sorbonnelaan 2, 3584 CA Utrecht, Netherlands
       \and
       R.Turolla \at
       Universit\'a di Padova,  via Marzolo 8, I-35131 Padova, Italy  
}

\date{Received: date / Accepted: date}
% The correct dates will be entered by the editor

\maketitle

\begin{abstract}
In the  last years, optical  studies of Isolated Neutron  Stars (INSs)
have expanded  from the more classical rotation-powered  ones to other
categories,  like the  Anomalous  X-ray Pulsars  (AXPs)  and the  Soft
Gamma-ray  Repeaters (SGRs),  which  make  up the  class  of the  {\em
magnetars}, the radio-quiet INSs with X-ray thermal emission and, more
recently, the  enigmatic Compact  Central Objects (CCOs)  in supernova
remnants.  Apart  from 10 rotation-powered pulsars,  so far optical/IR
counterparts have been found for 5  {\em magnetars} and for 4 INSs. In
this work we present some of the latest observational results obtained
from optical/IR observations of different types of INSs.

\keywords{Neutron Stars \and Optical \and Infrared}
\PACS{97.60.Gb \and 97.60.Jd}
\end{abstract}

\section{Introduction}
\label{intro}

Being   the   first   discovered   Isolated  Neutron   Stars   (INSs),
rotation-powered pulsars (RPPs) were also the first ones identified in
the optical. Recent summaries of  the RPPs optical observations can be
found  in  Mignani et  al.   (2004)  and  Mignani (2006).   After  the
spectacular results  of the 1990s, which  yielded to seven  of the ten
present RPP  identifications thanks to  the ESO \ntt\ (Mignani  et al.
2000a) and to  the \hst\ telescopes (Mignani et  al.  2000b), only PSR
J0437-4715 (Kargaltsev  et al.   2004) has been  added to  the record,
despite several attempts carried out after the advent of the ESO \vlt\
(e.g. Mignani et al.  1999,  2003, 2005; Mignani \& Becker 2004).  The
optical emission properties of RPPs  depend on the age, with the young
ones featuring purely magnetospheric  spectra and the middle-aged ones
featuring  composite  spectra  with  an additional  thermal  component
arising from the  cooling neutron star surface. For  older objects the
situation  is less  clear although  there is  evidence for  a dominant
magnetospheric emission (Mignani et  al.  2002; Zharikov et al. 2004),
while  only the  very old  PSR  J0437-4715 features  a purely  thermal
emission (Kargaltsev  et al. 2004).   \\ Multi-wavelength observations
carried out in  the last decades have unveiled  the existence of other
groups  of INSs,  most  of  them radio-quiet,  which  have been  later
studied  in   the  optical/IR.   \rosat\  observations   lead  to  the
identification  of  seven nearby  ($\le  300$  pc)  INSs dubbed  ``The
Magnificent  Seven'' (S.   Popov) with  purely thermal  X-ray emission
(Haberl,  this  conference).   Being  no unanimous  consensus  on  the
acronime to  use (J.  Tr\"umper, this  conference) from now  on I will
personally  refer to  these objects  as X-ray  Thermal  INSs (XTINSs).
Four XTINS have optical counterparts, with the identification of three
of them secured via proper motion measures.  Their inferred velocities
have also  allowed to rule out  surface heating from  ISM accretion as
the source of the thermal X-ray emission in favour of heating from the
cooling  neutron star  core.   The XTINS  optical  emission is  mostly
thermal and exceeds the extrapolation  of the soft X-ray spectrum by a
factor  $\sim 10$, which  suggests that  it arises  from a  cooler and
larger area on the neutron star  surface with respect to the X-ray one
(e.g.  Mignani et al.   2004).  Other peculiar INSs discovered through
their X-ray/$\gamma$-ray emission are the  AXPs and the SGRs which are
bealived  to  be  {\em  magnetars}, neutron  stars  with  hyper-strong
magnetic  fields  ($\sim  10^{14-15}$  G).   Out of  the  twelve  {\em
magnetars} so far identified (Woods  \& Thompson 2004), only four have
been  observed  in  the  optical/IR.   Very little  is  known  on  the
optical/IR spectra  of the {\em  magnetars}, apart from the  fact that
they  flatten with  respect to  the  extrapolation of  the soft  X-ray
spectrum.  This flattening  can be taken as an  indication of either a
turnover in the magnetar spectrum  or of the presence of an additional
emitting source (e.g.  an X-ray irradiated fallback disk).  Other very
enigmatic, supposedly  isolated, neutron stars are  the so-called CCOs
in SNRs (Pavlov et al.  2004).   Out of the seven CCOs known, only two
have proposed  optical/IR counterparts, classified as  low-mass $K$ or
$M$  stars (Sanwal et  al.  2002;  Pavlov et  al.  2004).   This would
suggest  that CCOs  are  indeed binary  rather  than isolated  neutron
stars.  Last  entry in the INS  family are the  newly discovered Rapid
Radio Transients or RRATs (Gaensler et al., this conference) for which
optical/IR  follow ups  have just  started.  \\  In the  following, we
present some  of the most  recent observational results  obtained with
the \vlt\  and the \hst\ in  the optical/IR studies of  some groups of
INSs, i.e.   RPPs, XTINSs and  CCOs.  For the {\em  magnetars}, recent
results are reported by Israel et al.  (this conference).

\section{Optical Observations of Rotation-Powered Pulsars}
\label{sec:1}
%and \cite{Ref1}

\subsection{The PSR J0537-6910 optical counterpart}
\begin{figure}
\centering
  \includegraphics[width=8.0cm]{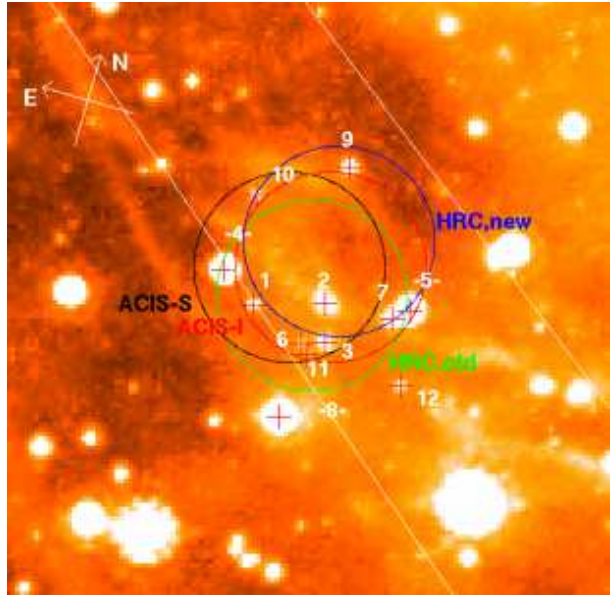}
\caption{\hst/\acs image  of the field of  PSR~J0537$-$6910 taken with
the  $814W$  filter with  candidates  labelled.  The  circles are  the
available \cxo\  positions of  the pulsar (Mignani  et al.  2005). The
projected orientation and width of the \stis/\nuv\ 52''$\times$2''slit
are shown.}
\label{psrj0537_acs}       % Give a unique label
\end{figure}

PSR~J0537$-$6910  is an  X-ray pulsar  (16  ms) in  the LMC  supernova
remnant  N157B.  The  measured  period derivative  $\dot  P \approx  5
\times 10^{-14}$s~s$^{-1}$  (Marshall et  al.\ 1998; Cusumano  et al.\
1998) implies a spin  down age of $ \sim  5\,000$~yrs and a rotational
energy loss of $\dot E  \approx 4.8 \times 10^{38}$ ergs~s$^{-1}$, the
highest  among RPPs.   As common  of young  pulsars,  PSR J0537$-$6910
exhibts  large glitches  (Middleditch  et al.\  2006)  and features  a
compact  pulsar-wind nebula  (Townsley  et al.\  2006).  Although  its
large  energy  output  makes  PSR~J0537$-$6910 a  natural  target  for
multi-wavelength observations,  it has  not yet been  detected outside
the  X-ray band.  In  radio it  is undetected  down to  $F_{1.4\, {\rm
GHz}} \sim 0.01$~mJy (Crawford et al.\ 2005), which implies that it is
significantly  fainter than  both the  Crab and  PSR  B0540-69.  First
exploratory optical  observations (Mignani et al.\  2000c; Gouiffes \&
\"Ogelman  2000; Butler  et al.\  2002)  also failed  to identify  the
pulsar counterpart, mainly owing to  the crowdness of the field.  More
recently,  deeper  high-resolution   observations  were  performed  by
Mignani et al.  (2005) with  the \acsn\ (\acs) aboard \hst\, and three
most  likely  counterparts  were  selected within  the  revised  \cxo\
position on the base of  their spectral flux distributions and colors.
\\ Follow-up  timing observations  of the candidate  counterparts have
been performed  with the  \stisn\ (\stis) aboard  the \hst\,  with the
\nuv\  (Mignani  at al.   2006a).   The  instrument  was used  in  its
spectroscopic configuration  with the PRISM  disperser (1460-3270 \AA)
and in  TIME-TAG mode (125 $\mu$  s time resolution).   The target was
observed  for five Continuous  Viewing Zone  orbits, yielding  a total
integration time of 25\,200 s.  All objects of Mignani et al.  (2005),
with the exception of 1, 4 and 8, fall in the 52'' $\times$ 2'' slit -
see Figure \ref{psrj0537_acs}.

\noindent
Unfortunately, two  objects only  are detected in  the two-dimensional
spectrum, of which  only one (object 5) is one  of the candidates. The
timing  analysis  does  not  reveil  evidence for  pulsations  at  the
expected  period, which  definetly rules  out object  5.  At  the same
time, the extinction corrected  [$E(B-V)=0.32$, Mignani et al. (2005)]
near-UV  flux upper limit  ($Log F_{\nu}  \sim -28.97$  ergs cm$^{-2}$
s$^{-1}$ Hz$^{-1}$) on the other candidates makes it unlikely that any
of them be the pulsar counterpart,  unless it has a very red spectrum.
The optical counterpart of the more and more elusive PSR J0537-6910 is
thus still unidentified.  This implies that, as in the radio band, PSR
J0537-6910  is intrinsically  fainter  than the  Crab  pulsar and  PSR
B0540-69.  This suggests that the optical luminosity of RPPs decreases
very fast, a scenario so far based only on the Vela pulsar case, which
is about  10 times older  than the Crab  but four orders  of magnitude
fainter.

\subsection{The optical polarization of the Vela Pulsar}
 Besides the radio band, optical polarimetric observations of RPPs and
of  their  synchrotron  nebulae  are  uniquely able  to  provide  deep
insights into the highly  magnetized relativistic environment of young
rotating neutron  stars.  Being the  first and the brightest  ($V \sim
16.5$) RPP  detected in the optical, polarization  measures were first
obtained for the Crab pulsar soon after its identification (Wampler et
al.  1969).   However, despite the substantial increase  in the number
of optically identified RPPs, the Crab is still the only one which has
both precise  and repeated polarization  measures (e.g.  Smith  et al.
1988).   Recently, Wagner \&  Seifert (2000)  performed phase-averaged
polarization observations of other  three young pulsars with the \vlt.
For the  Crab ``twin'' PSR B0540-69  ($V \sim 22.5$ )  they reported a
polarization of  $\approx$ 5\% (with no quoted  error bars), certainly
contaminated by  the contribution  of the surrounding  compact ($\sim$
4''  diameter) pulsar-wind nebula  (Caraveo et  al.~ 2001a).   For PSR
B1509-58 the value of the  optical polarization is also very uncertain
as the  newly proposed counterpart is  hidden in the PSF  wings of the
Caraveo et  al.  (1994) original  candidate ($V=22$).  Thus,  both PSF
subtraction  problems   and  the  object   faintness  ($R=26$),  whose
existance was never independently  confirmed so far, make the reported
polarization measure ($\approx 10$ \%, also quoted with no error bars)
tentative.  A polarization measurement of 8.5 $\pm$ 0.8 \% was finally
reported for  the Vela  pulsar ($V  \sim 23.6$).  \\  In order  to add
information on the polarizaton properties  of Vela, e.g.  the angle of
maximum polarization,  we have undergone  a careful reanalysis  of the
data set used  by Wagner \& Seifert.  The  details of the observations
and data reduction are described elsewhere (Mignani et al. 2006b).  We
obtained  a  polarization of  $9.4  \%  \pm  4 \%$,  qualitatively  in
agreement with  the one of  Wagner \& Seifert  but with a  much larger
error.  This is justified by the  fact that, owing to the faintness of
the target,  the uncertainty  on the background  subtraction dominates
the photometric errors on the  polarized fluxes, and ultimately on the
Stokes  parameters. We  are  thus  akin to  conclude  that this  large
difference  is ascribed  to an  error underestimation  on  their side,
likely to be  related to the neglection of  the background subtraction
contribution.  Thus, the value of the optical polarization of the Vela
pulsar, contrary to previous claims, is still uncertain.  We have also
computed  the angle  on  the plane  of  the sky  corresponding to  the
direction of  maximum polarization.  Interestingly, we  found that its
value ($\theta = 145^\circ \pm 14.7^\circ$) is compatible, perspective
wise,  with the  axis of  the X-ray  torus and  jet observed  by \cxo\
(Pavlov  et al.   2001) and  with  the pulsar's  proper motion  vector
(Caraveo  et al.   2001b).   Although we  can  not rule  out a  chance
coincidence,  it  is tentalizing  to  speculate  about the  alignement
between the  polarization direction  and the axis  of simmetry  of the
X-ray structures  as a tracer  of the connection between  the pulsar's
magnetospheric  activity and  its interactions  with  the environment.
More precise  measures of  the pulsar maximum  polarization direction,
possibly supported  by still to come polarization  measures in X-rays,
will  hopefully  provide  a  more  robust  observational  grounds  for
theoretical speculations.

\section{Optical/IR Observations of X-ray Thermal Isolated Neutron Stars}
\label{sec:2}

\subsection{The proper motion of RXJ 1605.3+3249}

\begin{figure}
\centering \includegraphics[width=8.0cm]{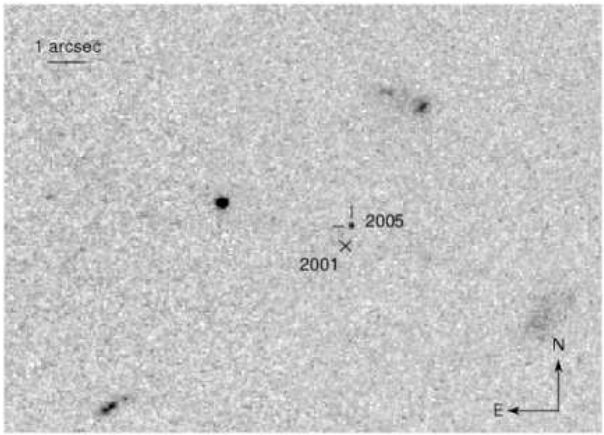}
\caption{\hst/\acs  image of  RX  J1605.3+3249 taken  in January  2005
(Zane et al.  2006). The counterpart is marked by  the two ticks while
the cross indicates its position measured in the 2001 \hst\stis\ image
of Kaplan et al. (2003).}
\label{rxj1605_pm}       % Give a unique label
\end{figure}

\begin{figure}
\centering \includegraphics[width=6.0cm,angle=-90]{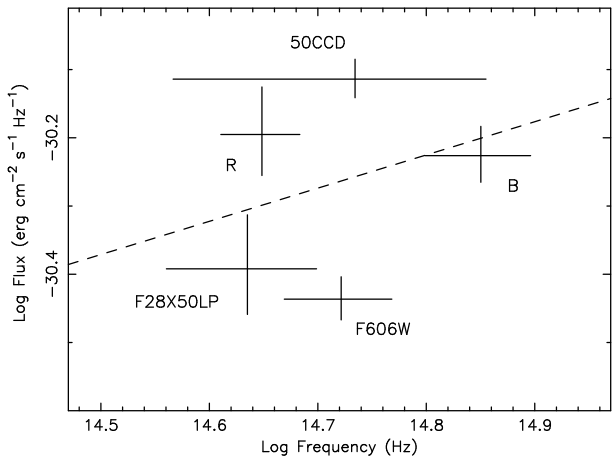}
\caption{Collection of the  currently available  (\hst \  and \sub)
optical  photometry  of  RX~J1605.3+3249,  corrected  for  a  reddening
$A_V=0.11$  (Zane  et al.  2006).  The dotted  line  represents  the best  fit
power-law ($\alpha =-0.5 \pm 0.5$).  }
\label{rxj1605_mc}       % Give a unique label
\end{figure}

An optical counterpart to RX  J1605.3+3249 was identified by Kaplan et
al. (2003)  in an  apparently blue object  detected with \hst\  at the
\cxo\ position.  As in the  case of other XTINS, e.g.  RX J1856.5-3754
(Van  Kerkwijk  \&  Kulkarni  2001)  and RX  J0720.4-3125  (Motch  et
al. 2003), the optical flux  of the candidate counterpart was found in
excess with respect to the Rayleigh-Jeans tail of the X-ray blackbody,
adding weight  to the proposed identification.  This  was confirmed by
the measure of the  object proper motion ($\mu=144.5\pm13.2$ mas/year,
position    angle    $\sim350.14^{\circ}    \pm   5.65^{\circ}$)    by
\cite{motch05}.  We have obtained  new observations of RX J1605.3+3249
with \acs\  to derive a more  accurate proper motion  measure (Zane et
al. 2006). By comparing the target position measured in our 2005 \acs\
image with the 2001 \stis\ one  of Kaplan et al.  (2003) we measured a
proper motion  $\mu = 155.0\pm3.1$ mas/year with  position angle $\sim
344^{\circ} \pm 1^{\circ}$  (see Figure \ref{rxj1605_pm}).  This value
confirms and  updates the one  of Motch et  al. (2005) and  settle the
optical identification of RX J1605.3+3249. Furthermore, it strenghtens
the  identification of the  neutron star  birth place  (for an  age of
$10^5$-$10^6$ years)  with the Sco~OB2 association,  also suspected to
be  the birth  place  of  other three  XTINSs.   The \acs\  photometry
(filter $606W$) has been compared with the one of Motch et al.  (2005)
and  Kaplan  et  al.    (2003)  to  characterize  the  source  optical
spectrum. While Kaplan et al. (2003), on the base of two \hst\ points,
suggested a blackbody,  Motch et al. (2005), on the  base of the \sub\
$B$ and $R$ points only,  claimed a non-thermal spectrum ($\alpha \sim
1.5$).  However,  by using  all available points  we can not  find any
statistically  acceptable fit (Figure  \ref{rxj1605_mc}), not  even by
excluding  the \acs\  point (Zane  et  al. 2006).   Thus, the  optical
spectrum of RX J1605+3249 is virtually unconstrained. New observations
taken  with the  same telescope  and  instrument set-up  to provide  a
consistent photometry are required.

\subsection{The search for the optical counterpart of RBS 1774}

\begin{figure}[h]
\centering \includegraphics[width=8.0cm,clip]{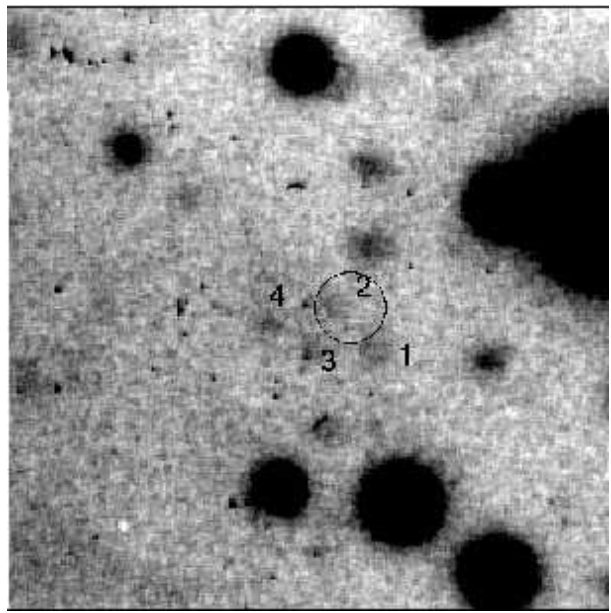}
\caption{\vlt/\fors\  $V$ band image  of the  RBS 1774 field  (one hour
integration  time) after  a gaussian  smoothing. The  objects detected
close to the \xmm\ 3'' error circle are labelled.}
\label{rbs1774}       % Give a unique label
\end{figure}

The  X-ray source 1  RXS 214303.7+065419  (aka RBS  1774) is  the last
entry in the XTINS family (Zampieri  et al. 2001) and one of the three
which still wait for an optical identification. As a part of dedicated
campaigns,  we have  carried out  \vlt\ observations  of  RBS 1774 with
\fors. Unfortunately, out  of the 8 hours observing  time ($B$ and $V$
bands) originally allocated in Service  Mode, only one hour in $V$ was
actually  executed.   Besides, the  quality  of  the observations  was
heavily affected by the very bad atmospheric conditions, with a seeing
costantly above 1.5''.  Figure  \ref{rbs1774} shows the $V$ band image
reduced through the  \fors\ pipeline, with the 3''  \xmm\ error circle
of RBS 1774 (Zampieri, private communication) overlaied. Although a few
objects (1-4) are detected, none of them can be considered a realistic
candidate counterpart to RBS 1774.  First  of all, they are at least a
factor 10  brighter than the optically identified  XTINS.  Then, after
comparing  our photometry with  the $B$  band one  of Komarava  et al.
(this conference)  obtained with  the \sub\, they  all turn out  to be
quite red, with  a $B-V > 0.5$.  This is  confirmed by their detection
in IR  \vlt\ images  (see next section).   No other object  has been
detected down to $V \sim 25.5$, which we set as the upper limit on the
RBS 1774 flux.

\subsection{Search for IR emission}

It  has  been noted  how  the  ``Magnificent  Seven'' show  intriguing
similarities with  the {\em magnetars}, which suggest  a possible link
between the two groups.  Their spin periods are similar (3--12 s) and,
through  the  observations  of  possible  cyclotron  X-ray  absorption
features,  magnetic field  of $B  \sim 6-7  \times 10^{13}$  have been
derived  in three  XTINSs.  Although  fainter than  those of  the {\em
magnetars}, they are one order of magnitude stronger than those of the
majority of radio  pulsars.  Also, for the two  XTINSs with a measured
period  derivative the  estimated X-ray  luminosities turn  out  to be
comparable  or larger  than the  inferred spin-down  energy.   In {\em
magnetars},  the  X-ray  luminosities  indeed exceed  their  spin-down
energy by at least 2  orders of magnitudes.  Finding more similarities
at other wavelengths  is certainly of unvaluable help  to strengthen a
possible  link   between  the  ``Magnificent  Seven''   and  the  {\em
magnetars}.  Both SGRs and AXPs are known to have peculiar IR spectra,
where   the  optical/IR   spectrum  flattens   with  respect   to  the
extrapolation of the X-ray spectra  (e.g.  Israel et al.  2003; Israel
et al.  2005).   Whatever the origin of this  spectral turnover, it is
interesting to  look for a similar  beahavior in the  XTINSs.  To this
aim, the best targets are RXJ 0720.4-3125 and RXJ 1856-3754 since they
are  the only  one  with  a rather  accurate  characterization of  the
optical spectrum (Motch  et al.  2003; Kaplan et  al.  2003).  In both
cases, a steep decline in the IR is expected, so that it would be easy
to  pinpoint  a  spectral  flattening,   if  it  is  present.   \\  IR
observations of both RXJ 0720.4-3125 and RXJ 1856-3754, as well as for
the three  unidentified XTINSs RXJ 0420-5022, RXJ  0806-4122, RBS 1774
are available  in the ESO  archive.  The observations were  taken with
the  \vlt\  between May  2004  and  December  2005 using  the  \isaac\
instrument with  the $H$ band  filter.  Integration times  are varying
between  4000 and  6000 s,  split  in shorther  dithered exposures  to
enable  for sky  subtraction.  The  data were  retrieved from  the ESO
archive together with the closest in time associated calibrations, and
reduced.   For RXJ  0720.4-3125 and  RXJ  1856-3754 we  have used  the
coordinates of  their optical  counterparts, while for  RXJ 0420-5022,
RXJ 0806-4122 and  RBS 1774 we have used the  availabe \cxo\ and \xmm\
coordinates.   In all  cases, no  object  was detected  at the  target
position, with the only exception  of RBS 1774 where we identified the
objects already  detected in our \fors\  $V$ band image  (see \S 3.2).
We  derived  $H$ band  upper  limits  of 21.9$\pm$0.15,  22.1$\pm$0.1,
22.4$\pm$0.1,  21.6$\pm$0.2 and  21.7$\pm$0.2 for  RXJ  0420-5022, RXJ
0720-3125, RXJ  0806-4122, RXJ  1856-3754 and RBS  1774, respectively.
Unfortunately, these  upper limits are  not very compelling.   For RXJ
0720-3125 and RXJ 1856-3754, a  spectral flattening redward of the $R$
band  would imply  a $H$  band  magnitude of  $\ge 24$  and $\ge  23$,
respectively. Similar  expectation values can be assumed  also for RXJ
0420-5022, RXJ 0806-4122 and RBS 1774 assuming similar optical spectra
and  a   factor  $\sim  10$   optical  excess  with  respect   to  the
extrapolation of their soft X-ray spectra.  The derived constraints on
a putative IR spectral flattening are not very compelling to constrain
the presence  of a  fossil disk,  either. By using  the disk  model of
Perna et al.  (2000) we were able only to exclude a disk extending at,
or beyond, the light cylinder.  Deeper IR observations to be performed
with \naco\ at the \vlt\ will allow us to improve these results.

\section{Optical Observations of Compact Central Objects (CCOs)}
\label{sec:3}

\subsection{1E 1207-5209 in G296.5+10.0}

\begin{figure}
\centering \includegraphics[width=8.0cm,angle=0,clip]{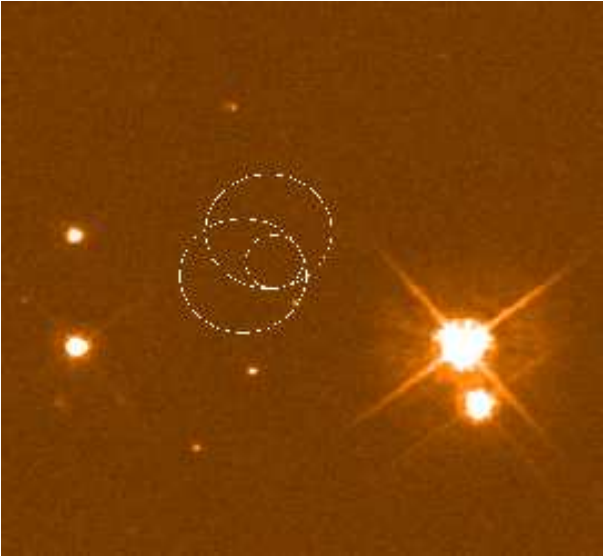}
\caption{\hst/\acs\ $814W$  filter image of  the 1E 1207-5209  field. The
upper  and lower  large  circles  ($1.5''$ radius)  are the  \xmm\
MOS1,MOS2 positions of the  source, respectively (DeLuca et al.  2004)
while  the  smaller  one  ($0.7''$  radius) is  the  revised  \cxo\
position.  Star A of Pavlov et  al. (2004) is visible at the
edge of the MOS2 error circle. }
\label{1207_acs}       % Give a unique label
\end{figure}

The CCO 1E 1207-5209 is one  of the very few which pulsates in X-rays,
with a period of 424 ms  (Zavlin et al.  2000) and a period derivative
$\dot{P}\sim1.4\times10^{-14}$ (Pavlov et al.  2002; Mereghetti et al.
2002; De Luca et al.   2004). Strangely, the characteristic age ($\sim
470,000$ years) is  about two orders of magnitude  higher that the age
($\sim 7000$  years) of the associated  SNR (Roger et  al.  1988).  In
absolute,  the  most  striking  peculiarity  of 1E  1207-5209  is  the
presence  of  three  (possibly  four)  X-ray  absorption  features  at
regularly spaced  energies.  These features were  interpreted in terms
of  electron or proton  cyclotron absorption  occurring in  a magnetic
field  of   $B\sim8\times10^{10}$  G  or   $B\sim1.6\times10^{14}$  G,
respectively  (Bignami et  al.  2003;  De Luca  et al.   2004).  These
values are about two orders  of magnitude lower/higher with respect to
the value  of the  magnetic field inferred  from the pulsar  spin down
($B_d\sim2\times10^{12}$  G).  Different hypotheses  to solve  the age
and B-field  discrepancies have been proposed,  including the possible
influence  of a  fossil disk  in the  neutron star  spin-down history.
Recently, Zavlin et al. (2004)  reported evidence for a non monotonous
spin evolution which could imply  that 1E 1207-5209 is either a strong
glitcher or it  is a binary.  To understand the  nature of the source,
very deep optical observations have been performed both with the \vlt\
and  with the  \hst.   De Luca  et  al.  (2004)  set  upper limits  of
$R\sim27.1$ and  $V\sim27.3$ on the optical brightness  of the source.
Soon after,  optical \hst and  IR \vlt\ observations unveiled  a faint
source (``star A'',  see Figure \ref{1207_acs}), apparently compatible
with  the  \cxo\ position.   The  source colors,  $m_{F555W}\sim26.8$,
$m_{F814W}\sim23.4$,   $J\sim21.7$,   $H\sim21.2$  and   $Ks\sim20.7$,
identify it  as a late $M$  star (Pavlov et al.   2004), thus implying
that  the 1E 1207-5209  is a  binary.  The  fact that  star A  was not
detected in  the optical images of  De Luca et al.   (2004) would also
imply that  it is variable.  \\  In order to  investigate the proposed
identification, we  have carefully reassessed the  \cxo\ astrometry of
1E1207-5209 and we have compared its  position with the one of star A.
Our  best   \cxo\  coordinates  are   $\alpha  (J2000)$=12$^h$  10$^m$
0.826$^s$,   $\delta  (J2000)$=  -52$^\circ$   26'  28.43''   with  an
associated uncertainty of $0.6''$.  The \hst/\acs\ images of the field
have been  retrieved from the  ESO archive after  on-the-fly reduction
and recalibration.  Single exposures have been combined using the {\it
multidrizzle}  task   in  IRAF,  which  also  corrects   for  the  CCD
geometrical distorsions.  For each final image we have then recomputed
the astrometry using  as a reference a number  of stars extracted from
the GSC2 catalogue.  The final  error of the target position is 0.7'',
inclusive  of the  accuracy  of our  astrometry  (0.17'').  The  \cxo\
position  is shown in  Figure \ref{1207_acs},  overlaied on  the \acs\
image  taken with  the $814W$  filter.  The  \cxo\ error  circle falls
within  the  intersection  of  the  MOS1  and  MOS2  ones  and  it  is
significantly offset  from the position of  the candidate counterpart,
which is right at the edge of the MOS1 error circle.  We thus conclude
that  star  A is  unlikely  to be  the  counterpart  to 1E  1207-5209,
although  the ultimate  piece of  evidence should  be obtained  by its
proper motion measurement, now in progress with the \hst.

\subsection{CXO J085201.4-461753 in Vela Jr}

\begin{figure}
\centering \includegraphics[width=8.0cm]{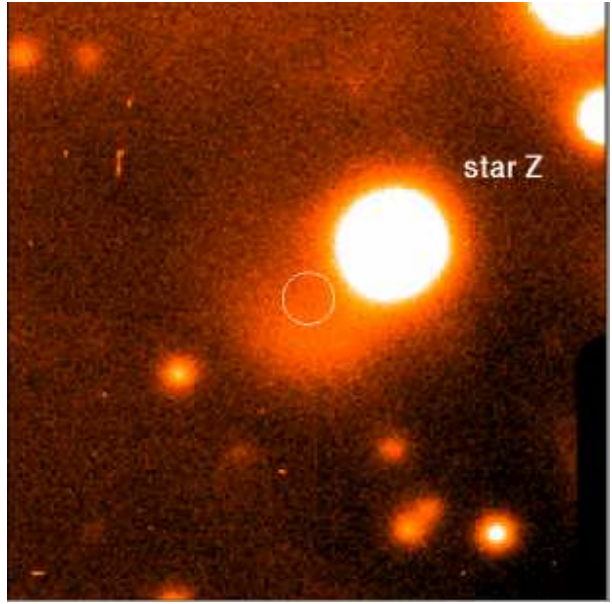}
\caption{$17'' \times 17''$ section of the \vlt/\fors\  $R$ band image
of the  Vela Jr. CCO. The circle  ($0.6''$ radius) corresponds
to the \cxo\ position uncertainty.  Star Z of Pavlov et al. (2001) is labelled. }
\label{velajr_fors}       % Give a unique label
\end{figure}

Vela  Jr. (G266.1-1.2) is  a very  young (a  few thousands  years) and
relatively nearby  ($\leq$1 kpc)  supernova remnant discovered  in the
ROSAT All Sky  Survey (Aschenbach 1998). The CCO in  Vela Jr was first
studied with \asca\  and \sax\ (Mereghetti 2001) and  later with \cxo\
which  also provided its  sub-arcsec position  (Pavlov et  al.  2001).
The  CXO  J085201.4-461753  X-ray   emission  is  characterized  by  a
thermal-like  spectrum,  as  in   other  CCOs,  with  no  evidence  of
pulsations (Kargaltsev  et al.  2002).  First  optical observations of
CXO J085201.4-461753 were presented by Pellizzoni et al.  (2002) using
archived  $B$  and  $R$  observations  taken  with  the  ESO/MPG  2.2m
telescope.  Although  no counterpart was  detected down to  $B=23$ and
$R=22$, the  digitized $H_{\alpha}$ plates  taken with the  UK Schmidt
telescope unveiled  the presence of an extended  emission blob ($\sim$
6''  diameter)  which  was  interpreted  as a  bow-shock  nebula  seen
face-on.  We  have performed deeper  observations of the Vela  Jr. CCO
with  \fors\ at the  \vlt.  To  minimize the  light pollution  from an
object (``star  Z'' of Pavlov et  al. 2001) located  $\sim$ 1.5'' away
from our target, we split the  integration time in 20 exposures of 260
s each.   In order  to achieve the  best possible  spatial resolution,
\fors\ was used in its High Resolution mode with a corresponding pixel
size of 0.1''.   A very bright star located 40''  away from the target
was  masked  using  the  \fors\  occulting  bars.   Observations  were
collected  with good seeing  ($\sim 0.9''$)  and airmass  ($\sim 1.3$)
conditions.  A $17'' \times 17''$ zoom of the \fors\ $R$ band image of
the  field   is  shown  in   Figure\ref{velajr_fors},  after  pipeline
reduction and  average combination of the single  exposures.  While no
point-like source appears at the \cxo\ position down to $R \sim 26$, a
compact optical nebula is detected.  We exclude that this nebula is an
artifact due to a PSF anomaly in star Z, to a defect in the image flat
fielding or to any instrumental  effect.  Both its position and extent
though are consistent with the one of the putative $H_{\alpha}$ nebula
seen by  Pellizzoni et al.  (2002), which clearly indicates  that they
are  the same  object.  Unfortunately,  the available  $B$  band upper
limit is  to shallow  to constraint the  nebula spectrum.  It  is thus
unclear wether it  is indeed a bow-shock or it  is some kind structure
similar to the pulsar-wind  nebulae seen around RPPs.  Follow-up \vlt\
observations, carried out at the  time of writing, will hopefully help
to unveil both the nature of this nebula and of the CCO.

\section{Infrared Observations of High Magnetic Field Pulsars}
\label{sec:4}

\subsection{PSR J1119-6127}

About 40 radio pulsars have  been detected by the Parkes Pulsar Survey
with magnetic fields larger than  $10^{13}$ G (Camilo et al. 2000). In
particular,  five of  them have  magnetar-like magnetic  fields larger
than the  quantum critical field $B_c  = 4.33 \times  10^{13}$ G above
which radio emission  is expected to be suppressed,  meaning that they
are not  expected to  be radio radio  pulsars at all.   Despite having
such  high magnetic  fields, these  high-magnetic field  radio pulsars
(HBRPs) do not behave as {\em magnetars}. First of all, they are radio
pulsars, while pulsed  radio emission has been discovered  so far only
in the transient AXP XTE  J1810-197 (Camilo et al. 2006). Second, only
two  HBRPs,  PSR  J1119-6127  (Gonzalez  \& Safi-Harb  2003)  and  PSR
J1718-3718 (Kaspi  \& McLaughlin 2005)  have been detected  in X-rays,
with luminosities $L_X \sim 10^{32-33}$ ergs s$^{-1}$ lower than those
of the {\em  magnetars}. Finally, they do not  show bursting emission,
either in  X-rays or  in $\gamma$-rays, as  AXPs and SGRs  instead do.
These differences  might be explained  assuming, e.g., that  HBPSs are
dormant  transients,  that  their   lower  X-ray  luminosities  are  a
consequence of  their lower magnetic  fields, or simply  assuming that
different  evolutionary  paths or  stages  account  for the  different
phenomenologies.  Of  course, one possibility is that  these HBRPs are
not genuine  {\em magnetars}  because the spin-derived  magnetic field
values are  polluted by  the torques produced  by a fossil  disk.  The
possible existance  of fossil disks around INSs  has been demonstrated
by the recent \spitzer\ discovery of  a disk around the AXP 4U 0142+61
(Wang et al.  2006).  Thus, if HBRPs do have fossil disks, they should
be detectable through deep, high-resolution, IR observations.  To this
aim, we  have started a program  of IR observations of  HBRPs with the
\vlt.  Since  the IR luminosity  of a disk  scales with the  X-ray one
(Perna et al.  2000), our  primary candidates are those HBRPs detected
in X-rays.   \\ The  field of PSR  J1119-6127 was observed  in Service
Mode between  January and February  2006 with NAos-COnica  (\naco), an
adaptive  optics imager  and spectrometer  at the  \vlt.  In  order to
provide   the  best   combination  between   angular   resolution  and
sensitivity, \naco\ was operated in  its S27 mode with a corresponding
field  of view  of  $28''\times28''$  and a  pixel  scale of  0.027''.
Observations  were performed in  the $J$,  $H$ and  $K_s$ bands  for a
total integration  time of 2 hours  each, dithered and  split in short
exposures  of  55 s  for  sky  subtraction  requirements.  The  seeing
conditions ($\sim 0.6''$)  allowed for an optimal use  of the adaptive
optics.   The data  were reduced  independently using  the  ESO \naco\
pipeline and  procedures run under the \eclipse\  package.  The pulsar
is undetected in any of the three observing passbands down to limiting
magnitudes  of $\sim$24,  $\sim$23 and  $\sim$22 in  the $J$,  $H$ and
$K_s$  passbands,  respectively. We  have  then  compared these  upper
limits with  the disk models  of Perna et  al.  (2000).  With  the due
caution that  the data analysis is  still in progress,  the results of
our simulations do  not presently allow to rule out  the presence of a
disk extending down to the magnetospheric radius.  Further theoretical
and simulation work  will allows to better constrain  wether, and how,
the putative disk interact with the neutron star.

\begin{acknowledgements}
Roberto  Mignani  warmly  thanks  the  ESO/Chile  Scientific  Visitors
Programme  for  supporting  his  visit  at the  ESO  Santiago  Offices
(Vitacura), where  part of  this work was  finalised. A  special thank
goes to  A. Micol  (ST-ECF) for  the support in  the reduction  of the
\hst/\acs\ data. 
\end{acknowledgements}

% BibTeX users please use
%\bibliographystyle{spmpsci}
%\bibliography{}   % name your BibTeX data base

% Non-BibTeX users please use

\end{document}